\documentstyle[12pt,epsfig]{article}
\def\gsim{\mathrel{\rlap {\raise.5ex\hbox{$ > $}}
{\lower.5ex\hbox{$\sim$}}}}
\def\lsim{\mathrel{\rlap {\raise.5ex\hbox{$ < $}}
{\lower.5ex\hbox{$\sim$}}}}

\newcommand{\be}{\begin{equation}}
\newcommand{\ee}{\end{equation}}
\newcommand{\bea}{\begin{eqnarray}}

\newcommand{\eea}{\end{eqnarray}}

\newcommand{ \HM} {HEIDEL\-BERG--MOSCOW~}

\newcommand{\BB} {Big Bang }

\newcommand{\SCM} {Standard Cosmology Model }
\newcommand{\MDG} {Majorana--Diraco genesis }
\newcommand{\SM} {Standard Model }
\newcommand{\SUSY} {supersymmetry }

\newcommand{\GUT} {Grand Unified Theory }
\newcommand{\BG} {baryo--genesis }
\newcommand{\LG} {lepto--genesis }
\newcommand{\SUGRA} {supergravity }
\newcommand{  \SS} {superstring }

\newcommand{ \gs}  {Gran Sasso Underground Laboratory }
\newcommand{\HP}{Poincar\'e }
\def\gappeq{\mathrel{\rlap {\raise.5ex\hbox{$>$}}
{\lower.5ex\hbox{$\sim$}}}}

\def\lappeq{\mathrel{\rlap{\raise.5ex\hbox{$<$}}
{\lower.5ex\hbox{$\sim$}}}}

\begin{document}

\begin{titlepage}

\begin{flushright}
CERN-PH/2006-118\\
PNPI/1000/06 \\
LAPTH/1155/06\\
\end{flushright}

\vspace{0.1in}
\begin{flushright}
To our Friend Tatiana Faberg{\'e}\\
\end{flushright}

\vspace{0.1in}
\begin{flushright}
We can only find  new ideas about \\
our World and suggest ways of \\
checking them in experiments. \\
But whether they are right or \\
wrong - only GOD knows.\\
Bruno Pontecorvo, Dubna, 1981 \\
\end{flushright}

\vspace{0.1in}
\begin{centering}

{\Large {\bf  Geometry of the  Majorana Neutrino}}

\vspace{0.1in}

{\Large{\bf  and  New Symmetries}}

\vspace{0.2in}

\Large{G. G. Volkov}
\footnote {On leave from PNPI, Russia}

\date{28.07.2006 }

\begin{abstract}

Experimental observation of Majorana fermion matter
gives a new impetus to the
understanding of the Lorentz symmetry and its extension, the
geometrical properties of the ambient space-time structure,
matter--antimatter symmetry and some new ways to understand the \BG
 problem in cosmology.
Based on the primordial Majorana fermion matter assumption, we
discuss a possibility to solve the  {\BG} problem through the
the {\MDG} in which we have a chance to understand  creation  of
$Q_{\rm em}$ charge and its conservation in our $D=1+3$ Universe after
the \BB.  In the {\MDG} approach there appears a  possibility to 
check the idea of the proton and electron non-stability on the
very low energy scale. In particle physics and in our space-time
geometry, the Majorana nature of the neutrino can be related to new
types of symmetries which are lying beyond the  binary Cartan-Killing-Lie
algebras/superalgebras. This can just support a conjecture about
the non-completeness of the SM in terms of binary Cartan--Killing--Lie
symmetries/supersymmetries. As one of the very important
applications of such new ternary symmetries could be related with
explanation of the nature of the three families and three colour
symmetry. Based on the {\MDG} the Majorana neutrino  can directly
indicate the existence of a new extra-dimensional geometry
(new cycle) and thanks  to new ternary  space-time symmetries, 
could lead at high
energies to the unextraordinary phenomenological consequences.

\end{abstract}
\end{centering}
\end{titlepage}

\vspace{0.5cm}

\indent

 We would like to discuss how the new space-time properties
of the neutrino, namely its  Majorana properties, may be an indication
of extra uncompactified space-time dimensions. The existence of
the Majorana neutrino gives new ideas about the possible composite
structure and form of electromagnetism in our Universe through the
relation between Dirac and Majorana fermions.

In this line of thought the experimental observation of the Majorana
 nature of the neutrino by the  \HM experiment  $0\nu\beta\beta$ decay
\cite{KK02,KK04-PL,KK04-NIMA} may have very important significance
for our attempts to build a theory that encompasses the Standard
Model.

Firstly, one can say that the discovery   of $0\nu\beta\beta$
decay is the indication for Majorana fermionic  matter in our
Universe as was expected not just in some extension of the \SM,
such as  $L--R$ models or \GUT, but also in all \SUSY, \SUGRA, {\SS}
approaches.

Secondly, the Majorana fermions can have a much  closer link with
the geometrical properties of our space-time and, and  most
important, the Majorana neutrino can give the first evidence  for
new uncompactified extra-dimensions space with $D \geq 1+3$ and,
correspondingly, the chance   to  observe the new space-time and internal
symmetries.

Extra-dimensional cosmology offers new possibilities to solve the
\BG problem  which,  correspondingly,  can give a  new universal
mechanism of the proton and electron non-stability. By
universality we mean that proton and electron decays are related
with the same dynamics. Also this new dynamics can change our
representations of  dark-matter and dark-energy problems.

However, extra uncompactified dimensions make quantum field
theories much less comfortable, since the power counting is worse \cite{BPHZ}
A possible way out is to suppose that the propagator is more
convergent than $1/p^2$,  such a behaviour can be obtained if we
consider, instead of binary  symmetry algebra, algebras with
higher order relations (\emph{that is, instead of binary
operations such as addition or product of 2 elements, one starts with
composition laws involving at least n elements of the considered
algebra, n-ary algebras}). For instance ternary symmetry could  be
related with membrane dynamics.

These  exotic symmetries can be related to algebras which are
based on  the generalization of the Lie binary commutation
relation, $[A,B]_2=AB -BA$, by the ternary  commutation relations:
$[A,B,C]_3=ABC+BCA+CAB-BAC-ACB-CBA$; they have already been
discussed in physics \cite{Nambu, Zuber, Kern, Dubois, Abramov,  Volkov2}, in
algebra \cite{Filippov}, in $CY_3$ geometry \cite{AENV, Volkov},
\cite{LSVV}, respectively. In \cite{DV} the  ternary symmetries
have been found through the ternary norm-division algebras of dimensions $3^n$,
{\it i.e.} ternary complex numbers ($n=1$), ternary "`quaternions" ($n=2$),
ternary "`octonions" ($n=3$).
The  ternary numbers directly give us the generalizations of the Abelian,
$su(3)$ and $g(2)$ binary algebras \cite{DV}.

The new symmetries beyond the well-known binary Lie
algebras/superalgebras could allow us  to build the renormalizable
theories for space-time geometry with dimension $D>4$. It seems
very plausible that using such ternary symmetries will offer a
real possibility to overcome the problems with quantization of
membranes and  could be a further progress  beyond string/{\SS}
theories. Also, these new ternary algebras could be related  with
some new {\SUSY} approach \cite{Abramov}. Getting the
renormalizable quantum field theories in $D\geq 4$ space-time we
access the point-like limits of the string and membrane
theories for some extra dimension $6 \geq D \geq 4$.

 The new,
so low, mass--energy scale ${\langle m\rangle}_{ee}=0.05-0.84\,{\rm eV}$
(the best value being
$0,39\, {\rm eV}$), $T_{1/2}^{0\nu}=(0.7--18.3) \times 10^{25}$ years
can indicate   the discovery of a  new space-time and of  internal
symmetries. We suppose that the ternary symmetry can be
responsible for
such low Majorana neutrino masses.
Combining this with the neutrino oscillation experiments
discovers an internal symmetry, leading to the mass relation
\begin{equation}
m_{\nu_e} =m_{\nu_{\mu}}=m_{\nu_{\tau}}, {\rm TERNARY \, RELATION}
\end{equation}
which could be the first  real indication of the existence of a
ternary symmetry whose nature is lying beyond the ordinary  Lie
algebra. This new ternary symmetry could shed light on the "`dark
symmetry"' of the SM:
\begin{equation}
N_{colour}=N_{family}=N_{dim}=3.
\end{equation}

This ternary mass relation  reminds the  binary Pauli--Luders relation
between masses of particles and antiparticle following in $D=3+1$ from
CPT invariance and $D=3+1$ Lorentz symmetry. We suppose that it could be followed
a ternary $D=6$ symmetry by generalized Lorentz symmetry and an analog of D=6
complete discrete symmetry in D=6. Thus we can conjecture such a relation just
for the particles living in D=6, like Majorana neutrinos, $\nu_e, \nu_{\mu}, \nu_{\tau}$.
In this case the mechanism of oscillation for neutrinos  can recall
the mechanism of oscillation $ K^0 --\bar K^0$, $B_d^0 --\bar B_d^0$,...
For this we should suggest a new interaction,
 which would mix the three  neutrino states.

The first consequence is related  an idea of complexity of
all Dirac spinor-fields in the \SM apart from Majorana neutrino
fields. The prediction of antimatter by Dirac in his relativistic
equation and the further experimental confirmation of his idea
were among  the exciting discoveries in particle physics. But this
discovery made a very intriguing puzzle for  cosmology, which was
not solved till now, 70-80  years after the  discovery of antimatter.

The discovery of the Majorana nature neutrino among the set of
all other \SM Dirac charged fermions  prompts another dynamics of
\BG, based on the composite dynamics of appearance of the fermion
Dirac matter from simple real fermions like Majorana neutrino
living in an extra-dimensional world?! The dynamics based on the
extra-dimensional geometry could be beyond Sakharov's three
conjectures. The fundamental conception of such an idea is related
attempts to find a common mechanism of the creation of
$Q_{\rm em}$ charge symmetry with \BG. We propose that such a
mechanism must use a duality between the $Q_{\rm em}$ conservation and
${\rm CPT}$ invariance:
\begin{eqnarray}
{\rm CPT}\,invariance \qquad \leftrightarrow \qquad (Q_{\rm em})\,
\,charge\,\, conservation,
\end{eqnarray}
{\it i.e.} the invariance of ${\rm CPT}$ in $D=4$ space-time
 means the conservation of the $Q_{\rm em}$
charge there, and opposite.

We could propose that primordially the Universe was neutral
$Q_{\rm em}=0$, more over, there were  absolutely no   charged
particles, in particularly no Dirac complex fermions. The Dirac
complex fermions appeared in $D=1+3$ ``brane world'' from real
high-dimensional  fermions as a result of a dynamics having
geometrical origin. In this case the part of a space-time where
such process took place must have the total charge equal zero.
We can propose that such   higher dimensional space-time world
has  a symmetry which is more fundamental than the symmetry
used in the \SM and \SCM, based on the Lie algebras and Lie
superalgebras. We already have some candidates for such
symmetries. They can be related with the  {\it n}-ary generalizations (
$n=3,4,...$) of Lie algebras and superalgebras. The geometry of such
space-time and  particle-states in this world  can be determined
by these  new universal symmetries, for example, based on the
unification of binary and ternary algebras. This could be a new
way to construct a realistic \GUT.
 Our proposal is  that the Majorana neutrino
could link with such neutral real fermions  and therefore
give us useful information about this high-dimensional world with
$D>4$. For terminology we can also  use the word `` brane world''
which means that the \SM matter is ``living'' in the three-
dimensional submanifold embedded into the higher-dimensional space
(a confinement)\cite{Rubakov}.

Usually, it was suggested that only gravitons can penetrate into
the bulk and only the neutrinos could interact with some bulk matter.
Based on the discovery of Majorana fermions we accept the more
general scenario in which the Majorana neutrinos can also be trapped
in the bulk and some exotic real Majorana-like fermions from the
bulk could interact weakly with Dirac fermions on the brane.

We can relate these real bulk fermions and neutrino to the
possibility of creation of Dirac fermions and appearance of the
$Q_{\rm em}$ charge symmetry in such world from some bulk real
fermions. Such dynamics could lead to the simultaneous production of baryon and
leptons  after \BB by the unique mechanism leading to
$Q_p+Q_e=0$. In such a scenario the   Majorana neutrinos
properties should allow it to have a ``memory''  about this
production-process?! We can study the physical models with extra dimensions
through the new symmetries, which could help us to
observe a duality between external  and internal dynamical
symmetries, {\it i.e.} to overcome the Coleman--Mandula no-go
theorem. One of the ways in this direction is related to the
search for generalizations of the principle of the  theory
of special  relativity, generalizations of the Klein--Gordon or
Dirac--Majorana equations \cite{Volkov2}. There were some attempts
to generalize  the  Dirac equation in
the frames of some ternary  Clifford algebra ( see for example, \cite{Abramov}).
The main question here is to get a "`relativistic"' invariant equation.
The ternary norm-division algebras \cite{DV} gives the way to construct a generalization
of the Lorentz symmetry \cite{DV} and , consequently, to get the invariant wave equations
for the particles living in $D>4$ space-time. Of course, the wave equations for the
new "`gauge"' fields will contain the derivatives of degree more than $2$.

Thus for our purpose we
can propose a correspondence between the $Q_{\rm em}$ charge
conservation and ${\rm CPT}$-invariance, which, in our opinion, could
have a geometrical origin in \HP duality \cite{Volkov2}. This correspondence
links  in $D=1+3$ space-time the matter--antimatter symmetry and
$Q_{\rm em}$ charge conservation. This proposal can be interpreted as
an attempt to find a correlation between external and internal
symmetries, overcoming by this the  Colleman-Mandula no-go theorem.
In our conjecture,  following to \HP duality, the internal
properties of the particle fields could have a correlation/link
with their space-time symmetry properties \cite{Volkov2}. Of course,  should be
valid the reverse sentence: one can study space-time through the
study of the particles properties.

In the beginning of the 90's, thanks  to \cite{Antoniadis} and later to
\cite{ADD}, the interest of the physics community was excited by
the suggestion of uncompactified new extra dimensions and their
link to  high energy physics near the \SM energy scale. The main
attraction of this approach is the possibility to reduce the mass
Planck scale from $10^{19} {\rm GeV}$ down to $\sim 1--10 $ {\rm TeV}, an
energy scale that can be explored experimentally in the  LHC. So,
a scenario with large or infinite uncompactified extra dimensions
can be more acceptable than one   based on a new physics
near $\sim 10^{19}$ {\rm GeV}.

 Actually, to study the
geometry of physical space-time with extra dimensions, one can
proceed by investigating its symmetries. To build a quantum field
theory in a space-time with extra dimensions ($D>4$), some of
which have a  status different from our usual 4 ones, it seems
inappropriate to use the usual the Lie--Killing--Cartan symmetries, as
compactification, mentioned above, has taught us. So there could
be just a chance to uncover new symmetries beyond the Lie-Cartan type.
We propose to reach  aim first  by  generalizing
binary Lie algebras/symmetries to $n$-ary algebras/symmetries
(\emph{that is, instead of binary operations like addition or
product of elements, one starts with composition laws involving at
least n elements of the considered algebra}).

A mechanism of geometrical origin of electromagnetic charge $Q_{em}$
and  Dirac complex  matter, baryon and lepton matter and
baryon--antibaryon (lepton--antilepton) asymmetry, proton--electron
non-stability  etc,  can be  called  \MDG and
could be considered as a further step in the  development
of \BG and \LG.
Note that   \BG-theory or \LG-theory demands to explain the baryon or
lepton asymmetry of the Universe
some exotic particles, lepto--quarks or right--handed neutrino,
correspondingly. The masses of such particles are very large:
$M_{lq} \sim 10^{18} GeV$ and $M_{RH}\sim 10^{12}--10^{14} GeV$,
respectively.
The \MDG is based on the existence of a new extra dimensional geometry
and can predict the proton/electron decays at the TeV mass scale,
which can be checked by  modern experiments (LHC).
Here one can find a similarity with the situation with
$M_{Planck}\sim 10^{19}GeV$ in $D=4$ and a possibility to diminish this
parameter using some extra dimensions \cite{Antoniadis}, \cite{ADD}.

The  idea of \MDG starts from Dirac/Majorana equations
and is related to the further attempts to solve
the baryon asymmetry of the Universe linking such question  with the  origin  $Q_{em}$
charge symmetry.
We would like to explore the idea of complexity of Dirac fermions
and propose that the Dirac fermions in $D=4$ have been produced
from real fermions of higher dimensions by  a complexification
or by a mechanism unknown to us.
There is one very remarkable fact

\begin{eqnarray}
|Q_p+Q_e|<10 ^{-21},
\end{eqnarray}
which can indicate that the proton (quarks) and the electron can have
a unique origin after the  \BB.

This proposal suggests the  existence  in high-dimensional space
of a class of real  fermions connected with our fermions through a new interaction,
which is based on a new symmetry beyond Lie,
such an interaction predicts a non-stability of the electrons or protons.
We can illustrate    scale of such an interaction,  taking for example
two extra dimensions. To make estimations we are in the situation in which
was  E.Fermi before the discovery the Yang-Mills interactions.
So we will follow the idea  he used for the  construction
of  the four-fermion weak theory with the coupling constant $G_{\rm}F$, of
 dimension
$[G_F]=[M^{-2}]$.
So we can start from the multi-fermion $D=6$ Fermi Lagrangian
and the corresponding Fermi constant $G_{F_S}$,
which could have  dimension:
\begin{equation}
[G_{\rm F_S}]= [M^{-p}], \qquad {\it p} =3,4,5,6,
\end{equation}
where {\it p} depends on the {\it n}-arity of a new interaction.

In  our opinion, this dimension of coupling constant corresponds
to a ternary gauge interaction, whose propagator could have a form
like $[P(q)_p+ M_s^p]^{-1}$, where $P(q)_p$ is a polynomial
of the transferred momentum $q$ of degree $p=3,4,...$.
 The form of such  a propagator corresponds
to a relativistic equation
of the ``gauge boson'' flying in $D=6$. We can  propose that such
a ``gauge boson''
will  membrane origin, but not of  string.
So, the gauge variant of the Fermi
$D=6$ interactions could have a  membrane origin and we can call them
membrane gauge bosons.
This is a completely new peculiarity of high-dimensional
interactions based on ternary gauge symmetries.
So for the tree-level calculations of the quark or charged-lepton decays into
neutral real bulk fermions $\nu_S$

\begin{eqnarray}
q \mapsto   m \, \nu_S, \qquad e^{\pm} \mapsto  m \, \nu_S,
\end{eqnarray}
taking the propagator of degree four   one can get the
following estimate for the partial width:

\begin{eqnarray}
\Gamma(e \rightarrow n S) \sim g_S^4 \cdot\frac{m_{e/q}^9}{M_S^8},
\end{eqnarray}
where $m_{e/q}$ is the electron mass
and $M_S$ is the membrane gauge boson mass.

To obtain the lower boundary for $M_S$
one can compare the partial width for electron decay with the lifetime 
of the muon
getting in $D_4$-Fermi interactions.
For the estimations one can take the  following
limit of the electron life-time:
$\tau_e \leq 10^{25}\,years \,\approx \pi \cdot 10^7 \cdot 10 ^{25}\, {\rm sec}$
This upper boundary gives the following estimate for

\begin{equation}
M_S \geq (\frac{g_2}{g_S})^{1/2}\cdot 10 \cdot M_W.
\end{equation}

From this, it is clear that the $M_S $ can be also in the
$\sim O(10) TeV$ region. This mass scale is consistent with the
upper boundary getting from the searching  the proton decay.
Note that in this approach  one can  study the proton decay problem
in $B$- physics, in $\mu-$ and $\tau- $ decays also!
There is also an interesting question about the spin structure of a new real
bulk fermions, $\nu_S$. Thank to the ternary symmetries they could have
the $1/3$ spin structure. On algebraic language it means that a map of the
internal ternary symmetry of  new fermions into the symmetry
of the ambient space-time is  a nontrivial triple covering
like in the case of the electron, its  spin $1/2$ structure is related with
the double covering maps:
$ SU(2) \mapsto SO(3)$ and/or  $ SL(2,C) \mapsto SO(3,1)$ \cite{Chevalley}.

We started our discussion from the Majorana neutrino and step by step
want to come to the idea of   existence of completely new physics
at the  energy scale $\sim O(10) TeV$. 
Firstly, this physics could be related to
the \BG problem  and proton/electron non-stability, which we also have
chances to observe in the $\mu-$, $\tau-$ lepton and $B$-meson rare decays.
Secondly, since  in our approach the Majorana neutrino ``lives'' in the 
higher-dimensional space-time describing by a new ternary symmetry generalized 
Lorentz symmetry we should also to change some fundamental principles 
of special theory of relativity, for example, the principle of maximal 
velocity. This one can check    by measuring the neutrino velocity
at high energies.If the energy scale $M_s\sim 10 TeV$, we can expect 
an increase  of the neutrino velocity $V/c-1\approx 10^{-4-5}$ at 
energies $E\sim 20--30 GeV$ \cite{LNGS}.
Thirdly, the Majorana-Diraco genezis gives us a chance to find the
mechanism of  creation of Dirac fermions and the $U(1)-$
electromagnetic vacuum of our Universe. This approach
suggests the existence of a new fermion matter with some new interactions 
based on the ternary gauge symmetry. Thus this matter can be 
the candidate for the dark matter.

To understand much deeper the role of Majorana neutrinos in the study the
ambient geometry of our  Universe
one can remind the role of photon and electron/antielectron in discovery of
the Lorentz symmetry and Minkowsky $D=3+1$ space-time.
At the end of the XIX century the lasting efforts
spent on the study of electromagnetism  had been finished by Maxwell
who wrote his fundamental equations \cite{Maxwell}
giving the  theory of  light.
Lorentz found that the Maxwell equations are satisfied  by the
symmetry, which now bears  his name. \HP extended the Lorentz
symmetry by translations. Based on the \HP and Lorentz groups it
was conjectured by Minkowski that  our world is a 4-dimensional,
Riemannian homogeneous space-time geometrical object with the
structure described by the metrics $g_{\mu\nu}=diag (1,-1,-1,-1)$.
Thus, one can say in  short, that the study of
electromagnetism--light  extended our geometrical representations
to the  ambient  structure of our world from dimension $D=3$ to
dimension $ D=4$.

Later, the theory of the electron just confirmed this  conjecture.
The spin structure of the electron can be described by a spinor wave function
described by the $SU(2)$ group symmetry which is the double covering group
of the   $SO(3)$ group symmetry of  the  spatial part of our
space. The extension of the external SO(3) group to the  Lorentz
group $SO(1,3)$, with its double covering group $SL(2,C)$,
 gave some
additional degree of freedoms in the four-spinor dimensional
spinor wave function which was used by Dirac to discover the
antiparticles that were  nicely confirmed by experiments. The
theory of light and electron have been unified in quantum
electrodynamics, based on the \HP  /Lorentz  symmetry and
$U(1)_{em}$ gauge symmetry acting in the Minkowski  D=1+3
space-time. One of the fundamental consequences of this quantum
theory is the CPT theorem, which tells us that these two families
of matter and antimatter should have some ``degenerate''
properties, {\it i.e. } the masses and life-times  of particle and
antiparticle must coincide.
 The ternary  mass relation $m_{\nu_1}=m_{\nu_2}=m_{\nu_3}$
can be expained in analogy with two
``generations'' of particle and antiparticle in the binary Lorentz group,
where the CPT theorem gives the exact relations between their masses:
$m=\bar m$. If the origin of the three generations really is related
to ternary symmetry,  similar discrete symmetries could give such
ternary mass relation for the masses of the  three Majorana neutrinos
living in $D=6$. Since the Dirac fermions are
living in $D=4$ the ternary mass relation cannot be valid more.
 The gauge invariance principle gave us  the complete picture of
the quantum theory of light
which can be  formulated in the language of  the external and internal
symmetry \HP duality:
\begin{eqnarray}
SO(3,1) \leftrightarrow  U(1)^{\rm em}
\end{eqnarray}
Thus if we believe that we could understand the geometrical origin of $U(1_{\rm em})$
symmetry we must do the next step: to understand the geometrical origin of the  $SM$,
{\it i.e.} to understand the geometrical origin of the
$SU(3^c)$, $SU(2_{\rm EW})\times U(1_Y)$ gauge symmetries and, consequently,
to explain why the
$SU(3^c)$- symmetry  is exact and the $SU(2_{\rm EW})\times U(1_Y)$ symmetry  is broken?!
The last comment is related with an alternative  way to the   Higgs mechanism
of breaking the $SU(2) \times U(1)$ gauge symmetry to the $U(1_{\rm em})$ symmetry.

Our conjecture is that if the Majorana neutrinos living
in the bulk can be  described by a new ternary  symmetry which is
a generalization of the Lorentz symmetry. If the extra-dimensional
space-time produce  a new geometrical cycle the maximum speed limit for
the high energetic  neutrinos could be
different from  the  velocity of the light  and could  be
connected with the new fundamental constant of the second cycle,
{\it i.e.} $C_2 \geq c_{\nu} \geq C_{1} \equiv c$
\cite{Volkov2, AV, LNGS}.
Thus Majorana neutrinos can help us to go beyond the Lorentz symmetry
and to see  a new world through the new symmetries and new
 "`gauge" interactions
such  as the  membrane gauge bosons that can
give an origin of a new ``light''related with a new Abelian gauge
symmetry, which would be  called by
``neutrino light'' or ``dark matter light'' \cite{nlight}:

\section{Acknowledgements}
I would  like to thank particularly  Tatiana Faberg{\'e}
for her wonderfull sense of the permanent help during 
the long period of working on this article.

The idea to write the article about the Majorana neutrinos
was initiated by H.V. Klapdor-Kleingrothaus and I. Krivosheina, 
who invited me to the Gran-Sasso Laboratory  on summer 2005,
where I have got nice possibilities to
study the intriguing results of the \HM experiment 
\cite{KK02,KK04-PL, KK04-NIMA}.
 During seven last  years we already discussed  the possible 
role of Majorana neutrino in  \MDG.  
The Majorana nature of neutrino 
 is playing the crucial role in this scheme.
This idea was closely related with the possible new 
geometrical properties ($D>4$) of  Majorana neutrino to check 
which we  suggested in measuring
the neutrino velocity at high energies \cite{AV}.
The progress what made in the construction of the new
neutrino detectors
in  \gs gives us the  independent possibilities  to check is  neutrino
of Majorana nature or not \cite{AV}.
This question is dramatically difficult both in  theory and
in  experiment.(The neutrino  short base experiment
to study if the   neutrino  is a tachyon or not, was done in FNAL in
the 1977.)
The other possibility to check the Majorana nature of neutrino
can be related with the proton decay problem. In the scheme of \MDG the
proton non-stability problem could be solved also in the $e-$, $\mu-$,
$\tau-$ lepton
and $b-$ meson  rare decay experiments. 
The Majorana nature of the neutrino gave  a lot of new and non-ordinary
ideas  for the probing of
 which I would like to express my thanks
to L. Alvarez-Gaum{\'e}, U. Aglietti,  V. Ammosov, I. Antoniadis,  P. Aurenche,
M. Baldo-Ceolin, G. B{\'e}langer,
G. Costa, P. Chankowski, A. De Rujula, A. Dubrovskiy, J. Ellis, L. Fellin,
H. Klapdor-Kleingrothaus, I. Krivosheina,  
L. Lipatov, A. Liparteliani, A. Masiero,
R. Obergfeld, M. Ruggier, A. Sabio-Vera, J-B. Zuber.
It is pleasure for me to thank Z.Berezhiani  and the participants 
of the X Summer Institute 2005 and XI Summer Institute 2006  
 in Gran-Sasso for the nice possibilities to discuss the  ideas of this work. 

 I also thank very much   G. Girardi for reading manuscript
and many comments.

\end{document}